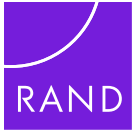

Expert Insights
PERSPECTIVE ON A TIMELY POLICY ISSUE

OLIVER GUEST, KEVIN WEI

# Bridging the Artificial Intelligence Governance Gap

The United States' and China's Divergent Approaches to Governing General-Purpose Artificial Intelligence

December 2024

For more information on this publication, visit **www.rand.org/t/PEA3703-1**.

### About RAND

RAND is a research organization that develops solutions to public policy challenges to help make communities throughout the world safer and more secure, healthier and more prosperous. RAND is nonprofit, nonpartisan, and committed to the public interest. To learn more about RAND, visit www.rand.org.

### Research Integrity

Our mission to help improve policy and decisionmaking through research and analysis is enabled through our core values of quality and objectivity and our unwavering commitment to the highest level of integrity and ethical behavior. To help ensure our research and analysis are rigorous, objective, and nonpartisan, we subject our research publications to a robust and exacting quality-assurance process; avoid both the appearance and reality of financial and other conflicts of interest through staff training, project screening, and a policy of mandatory disclosure; and pursue transparency in our research engagements through our commitment to the open publication of our research findings and recommendations, disclosure of the source of funding of published research, and policies to ensure intellectual independence. For more information, visit www.rand.org/about/research-integrity.

RAND's publications do not necessarily reflect the opinions of its research clients and sponsors.

# About This Paper

The United States and China are among the world's top players in the development of general-purpose artificial intelligence (GPAI) systems, and both are keen to lead in global artificial intelligence (AI) governance and development. A look at U.S. and Chinese policy landscapes reveals differences in how the two countries approach AI governance. Three areas of divergence are notable for policymakers: the focus of domestic AI regulation, key principles of domestic AI regulation, and approaches to implementing international AI governance. This work should be of interest to policymakers and researchers interested in AI governance in China and international AI governance.

## Technology and Security Policy Center

RAND Global and Emerging Risks is a division of RAND that delivers rigorous and objective public policy research on the most consequential challenges to civilization and global security. This work was undertaken by the division's Technology and Security Policy Center, which explores how high-consequence, dual-use technologies change the global competition and threat environment, then develops policy and technology options to advance the security of the United States, its allies and partners, and the world. For more information, contact tasp@rand.org.

## Funding

This research was, at its inception, independently initiated and conducted within the Technology and Security Policy Center using income from operations and gifts from philanthropic supporters, which have been made or recommended by DALHAP Investments Ltd., Effektiv Spenden, Ergo Impact, Founders Pledge, Fredrik Österberg, Good Ventures, Jaan Tallinn, Longview, Open Philanthropy, and Waking Up Foundation. This research was published through funding provided by DALHAP Investments Ltd., as recommended by Ergo Impact and Fathom. A complete list of donors and funders is available at www.rand.org/TASP. RAND donors and grantors have no influence over research findings or recommendations.

## Acknowledgments

We thank Michael Aird, Renan Araujo, Karson Elmgren, Brodi Kotilla, Ulysse Richard, Aris Richardson, Konrad Seifert, and Zoe Williams for their feedback. We also thank our reviewers, William C. Hannas and Bonny Lin. The people thanked here do not necessarily agree with the content of the paper and any errors are our own.

# Bridging the Artificial Intelligence Governance Gap: The United States' and China's Divergent Approaches to Governing General-Purpose Artificial Intelligence

The United States and China are among the world's top players in the development of advanced artificial intelligence (AI) systems, and both are keen to lead in global AI governance and development.[1] A look at U.S. and Chinese policy landscapes reveals differences in how the two countries approach AI governance. The following three areas of divergence are notable for policymakers:

- the focus of domestic AI regulation
- key principles of domestic AI regulation
- approaches to implementing international AI governance.

These differences in approach might reflect broader challenges for international cooperation between the United States and China on AI safety and security. As AI development continues, global conversation around AI, such as statements from various governments,[2] dialogues,[3] and reports authored by prominent researchers from both countries,[4] has warned of global safety and security challenges posed by general-purpose AI (GPAI) systems. GPAI systems, such as large language models, can be used in a wider variety of contexts compared with narrow or domain-specific AI systems.

---

[1] Ben Cottier, Tamay Besiroglu, and David Owen, "Who Is Leading in AI? An Analysis of Industry AI Research," arXiv, arXiv:2312.00043, November 24, 2023; Sihao Huang, "Beijing's Vision of AI Global Governance," *ChinaTalk* blog, October 23, 2023; John Thornhill, "Step Aside World, the U.S. Wants to Write the AI Rules," *Financial Times*, November 2, 2023.

[2] Government of the United Kingdom, "The Bletchley Declaration by Countries Attending the AI Safety Summit, 1–2 November 2023," policy paper, November 1, 2023.

[3] International Dialogues on AI Safety, "About," webpage, undated.

[4] Yoshua Bengio, Geoffrey Hinton, Andrew Yao, Dawn Song, Pieter Abbeel, Trevor Darrell, Yuval Noah Harari, Ya-Qin Zhang, Lan Xue, Shai Shalev-Shwartz, et al., "Managing Extreme AI Risks Amid Rapid Progress," *Science*, May 20, 2024; Bengio, Y., Mindermann, S., Privitera, D., Besiroglu, T., Bommasani, R., Casper, S., Choi, Y., Goldfarb, D., Heidari, H., Khalatbari, L., et al., *International Scientific Report on the Science of AI: Interim Report,* Government of the United Kingdom, May 2024; Matt Sheehan, "China's Views on AI Safety Are Changing—Quickly," Carnegie Endowment for International Peace, August 27, 2024b.

Cooperation between the United States and China might be needed to address these risks, and experts and policymakers from both sides have repeatedly called for such cooperation since the rapid development of GPAI systems in the late 2010s.[5] For example, experts have warned against developing AI systems that cross key redlines, such as the ability to autonomously deceive humans or to execute devastating cyberattacks. Cooperation could help mitigate competitive pressures to develop AI systems that are highly strategic but risk crossing redlines.[6]

Understanding areas of divergence will highlight areas in which additional consensus-building and discussion could be needed between the United States and China. Additionally, understanding these differences could help facilitate cooperation by encouraging interoperability and standardization. In cases in which U.S. and Chinese AI governance diverge, policymakers could also learn from their counterparts' efforts, even if they disagree with their counterparts' policy objectives.

## A Short History of U.S. and Chinese GPAI Governance

Since 2017, both the United States and China have demonstrated a willingness to govern GPAI.[7] Domestic policy is not a perfect proxy for international governance, but some strategic lessons might still be drawn from domestic policy in terms of governance priorities on the global stage. In the United States, Executive Order 14110 initiated risk assessment and reporting requirements for some GPAI systems.[8] Meanwhile, the Chinese Communist Party resolved in a 2024 plenary session to "institute oversight systems to ensure the safety of artificial intelligence"; an official statement from party leadership explicitly linked that decision to recent rapid progress in GPAI development.[9]

There have already been some moves toward cooperation between the two countries on reducing safety and security risks from GPAI.[10] Key examples include a 2023 AI Safety Summit held in the United Kingdom and its successors, two resolutions at the United Nations (UN) General Assembly, and the U.S.-China intergovernmental dialogue on AI.[11] Although some of these initiatives might be policy signaling rather than commitment, these early initiatives reflect at least an awareness of GPAI risks.

---

[5] Ryan Browne, "Google Deepmind Boss Hits Back at Meta AI Chief over 'Fearmongering' Claim," CNBC, October 31, 2023; Concordia AI, *State of AI Safety in China*, October 2023; Mariano-Florentino Cuéllar and Matt Sheehan, "AI Is Winning the AI Race," *Foreign Policy*, June 19, 2023; Henry A. Kissinger and Graham Allison, "The Path to AI Arms Control," *Foreign Affairs*, October 13, 2023; Karman Lucero, "Managing the Sino-American AI Race," *Project Syndicate*, August 9, 2024; Madhumita Murgia, "White House Science Chief Signals US-China Co-Operation on AI Safety," *Financial Times*, January 24, 2024; Joseph S. Nye, Jr., "U.S.–China Cooperation Remains Possible," *Project Syndicate*, May 6, 2024; Huw Roberts, "China's Ambitions for Global AI Governance," *East Asia Forum*, September 10, 2024; Sun Chenghao and Lie Yuan, "How Should China and the United States Promote Artificial Intelligence Dialogue and Cooperation?" ["中美应如何推动人工智能对话与合作"], *China US Focus*, February 6, 2024; Catherine Thorbecke, "Where the U.S. and China Can Find Common Ground on AI," Bloomberg, September 19, 2024.

[6] International Dialogues on AI Safety, statement at IDAIS-Beijing 2024, March 10–11, 2024; "Sabotage Evaluations for Frontier Models," *Anthropic*, October 18, 2024.

[7] "Is Xi Jinping an AI Doomer?" *The Economist*, August 25, 2024; Graham Webster, "What If 'AI Safety' Is No Longer a Constructive U.S.-China Topic?" *Here It Comes* blog, August 23, 2024.

[8] Executive Order 14110, "Safe, Secure, and Trustworthy Development and Use of Artificial Intelligence," Executive Office of the President, October 30, 2023.

[9] Central Committee of the Communist Party of China, resolution on Further Deepening Reform Comprehensively to Advance Chinese Modernization, July 18, 2024; Concordia AI, "What Does the Chinese Leadership Mean by 'Instituting Oversight Systems to Ensure the Safety of AI?'" *AI Safety in China* blog, August 2, 2024a.

[10] Nayan Chandra Mishra, "From Competition to Cooperation: Can US-China Engagement Overcome Geopolitical Barriers in AI Governance?" *Tech Policy Press*, September 23, 2024.

In 2023, the United States and China both took part in an AI Safety Summit in the United Kingdom; the summit resulted in the Bletchley Declaration on AI risks.[12] The declaration was signed by 29 jurisdictions, including the United States and China, and warns that "there is potential for serious, even catastrophic, harm, either deliberate or unintentional" from AI and international cooperation would be necessary to tackle these risks.[13] Both countries attended a subsequent summit in South Korea.[14] These summits have led to the International Scientific Report on the safety of advanced AI and commitments by some AI companies to specific risk-management practices.[15]

At the UN General Assembly, both countries have introduced resolutions that articulate principles about AI. The U.S.-led resolution, which China supported, focuses on "safe, secure and trustworthy" AI.[16] The Chinese-led resolution, which was in turn supported by the United States, is primarily about capacity-building, to the end of "harnessing the benefits of artificial intelligence, minimizing its risks, and accelerating innovation and progress toward the achievement of all 17 Sustainable Development Goals."[17] (However, China's stated desire to engage with other countries might be concerning in light of historic exports of surveillance-enabling technologies.[18])

Government representatives also met in May 2024 for a dialogue about risks associated with advanced AI. Readouts from both countries noted that such systems could be dangerous and that the dialogue had been constructive.[19] Neither nation made concrete commitments, but they did agree to a second meeting.[20]

These examples demonstrate some shared willingness to cooperate to reduce safety and security risks from GPAI. However, concrete initiatives for cooperation have been blocked by various factors, such as the need for technical mechanisms to verify formal agreements and the tense bilateral relationship between the two countries.[21] This relationship is often framed through the lens of technological competition, such as the United States using export controls to slow China's development of GPAI.[22]

---

[11] Government of the United Kingdom, 2023; UN General Assembly, resolution on Seizing the Opportunities of Safe, Secure and Trustworthy Artificial Intelligence Systems for Sustainable Development, Resolution A/78/L.49, March 11, 2024a; UN General Assembly, Resolution on Enhancing International Cooperation on Capacity-Building of Artificial Intelligence, Resolution A/78/L.86, June 25, 2024b; Richard Weitz, "China and the United States Begin Official AI Dialogue," *China US Focus*, June 14, 2024.

[12] Government of the United Kingdom, 2023.

[13] Government of the United Kingdom, 2023.

[14] Government of the United Kingdom, "AI Seoul Summit: Participants List (Governments and Organisations)," webpage, last updated June 13, 2024b.

[15] Bengio, Mindermann, et al., 2024; Government of the United Kingdom, "Historic First as Companies Spanning North America, Asia, Europe and Middle East Agree Safety Commitments on Development of AI," press release, May 21, 2024a.

[16] UN General Assembly, 2024a, p. 1.

[17] UN General Assembly, 2024b, p. 3.

[18] Martin Beraja, Andrew Kao, David Y. Yang, and Noam Yuchtman, "Exporting the Surveillance State via Trade in AI," Brookings Institution, July 24, 2023; Martin Beraja, David Y. Yang, and Noam Yuchtman, "China Is Exporting Its AI Surveillance State," *Project Syndicate*, July 24, 2024; Ministry of Foreign Affairs of the People's Republic of China, "AI Capacity-Building Action Plan for Good and for All," September 27, 2024b.

[19] "China's Readout of the First Sino-U.S. Intergovernmental Dialogue on AI," *Geopolitechs* blog, May 14, 2024; White House, "Statement from NSC Spokesperson Adrienne Watson on the U.S.-PRC Talks on AI Risk and Safety," May 15, 2024.

[20] "China and the U.S. Agreed to Hold the 2nd Intergovernmental Dialogue on AI," *Geopolitechs* blog, August 28, 2024.

Another barrier could be differences in approaches to AI governance between the United States and China. As noted previously, three areas of divergence in particular are important for policymakers to understand: differences in the focus of AI regulation, differences in key principles of AI regulation, and differences in approaches to implementing international AI governance.

## Differences in the Focus of AI Regulation

AI policy proposals in the United States have generally focused on AI models and systems; in contrast, Chinese AI policy typically has its roots in content regulation. Chinese content regulation could often be described as political censorship but also includes such goals as limiting content that could be used to harm others.

*AI models* are the core algorithms that process information and generate outputs; *AI systems* contain these models within the broader infrastructure that influences their operation, such as content filters, system prompts, and user interfaces.[23] These terms are key concepts in Executive Order 14110, which reflects their centrality in U.S. policy discussions; similar emphasis on AI models or systems can be found in Colorado state legislation and a vetoed California Senate Bill.[24]

U.S. policy proposals typically impose safety requirements on AI model developers. Many proposals include provisions that impose heightened scrutiny on models that are trained using more than a specified amount of computing power. This approach stems from the view that increases in training compute correlate to model capabilities and risks.[25] By focusing on models and systems, U.S. policymakers aim to regulate the inherent properties and capabilities of AI technologies.

In contrast, and in line with China's long-standing implementation of information controls,[26] AI governance in China has historically focused more on AI outputs (and sometimes data inputs that result in those outputs) than on the models or systems themselves. For example, a standard released in February 2024 by a national-level standardization body in China lays out exacting methods for evaluating the safety of model-generated content.[27] This focus can also be seen in two major regulations that apply to GPAI systems, both of which appear to primarily target actors that deploy models, rather than actors that develop them (although these can overlap).[28] Moreover, no official policy document references a compute threshold for AI models that is analogous to the one referenced in Executive Order 14110, although two academic-led drafts of an AI law do contain unspecified compute thresholds.[29] The United States does not have content regulations analogous to China's regulations.

---

[21] Miles Brundage, Shahar Avin, Jasmine Wang, Haydn Belfield, Gretchen Krueger, Gillian Hadfield, Heidy Khlaaf, Jingying Yang, Helen Toner, Ruth Fong, et al., "Toward Trustworthy AI Development: Mechanisms for Supporting Verifiable Claims," arXiv, arXiv:2004.07213, April 20, 2020; Yonadav Shavit, "What Does It Take to Catch a Chinchilla? Verifying Rules on Large-Scale Neural Network Training via Compute Monitoring," arXiv, arXiv: 2303.1134, May 30, 2023.

[22] Gregory C. Allen, "Choking Off China's Access to the Future of AI," Center for Strategic and International Studies, October 11, 2022; Stephen Cave and Seán S. ÓhÉigeartaigh, "An AI Race for Strategic Advantage: Rhetoric and Risks," AIES '18: Proceedings of the 2018 AAAI/ACM Conference on AI, Ethics, and Society, December 2018.

[23] Lee Sharkey, Clíodhna Ní Ghuidhir, Dan Braun, Jérémy Sheurer, Mikita Balesni, Lucius Bushnaq, Charlotte Stix, and Marius Hobbhahn, *A Causal Framework for AI Regulation and Auditing*, Apollo Research, November 2024.

[24] California State Senate, Safe and Secure Innovation for Frontier Artificial Intelligence Models Act, Senate Bill 1047, September 2, 2024; General Assembly of the State of Colorado, Concerning Consumer Protections in Interactions with Artificial Intelligence Systems, Senate Bill 24-205, May 17, 2024.

[25] Lennart Heim and Leonie Koessler, "Training Compute Thresholds: Features and Functions in AI Regulation," arXiv, arXiv:2405.10799, August 6, 2024.

[26] Qianer Liu, "China to Lay Down AI Rules with Emphasis on Content Control," *Financial Times*, July 10, 2023.

Chinese policymakers might be moving toward regulating AI models and systems in addition to or instead of content and outputs. Although policymakers might have been motivated by concerns about AI-generated content, Chinese algorithm registries have forced companies to disclose some information about their AI models.[30] The resolution from a 2024 Chinese Communist Party plenary session has also stated an intention to "institute oversight systems to ensure the safety of artificial intelligence."[31] It remains uncertain which risks Chinese regulations will ultimately address; it is also unclear how China will choose to scope and define GPAI or other advanced models for regulation—but this recent shift suggests increased policy alignment with U.S. approaches to AI regulation.

## Differences in Key Principles of AI Governance

Statements, such as the Bletchley Declaration on AI Risk and those from various Track II dialogues,[32] demonstrate many shared AI-governance priorities between the United States and China.[33] However, there are divergences driven by competing values on at least three principles: who should be able to develop and deploy GPAI, control over AI systems, and state power.[34]

---

[27] National Technical Committee 260 on Cybersecurity of Standardization Administration of China, *Basic Safety Requirements for Generative Artificial Intelligence Services* [生成式人工智能服务安全基本要求], trans. by Center for Security and Emerging Technology, February 2024.

[28] Cyberspace Administration of China, "Regulations on the In-Depth Synthesis Management of Internet Information Services" ["互联网信息服务深度合成管理规定"], December 11, 2022; Cyberspace Administration of China, "Interim Measures for the Management of Generative Artificial Intelligence Services" ["生成式人工智能服务管理暂行办法"], July 13, 2023.

[29] Zhang Linghan, Yang Jianjun, Cheng Ying, Zhao Jingwu, Han Xuzhi, Zheng Zhifeng, and Xu Xiaob, *Artificial Intelligence Law of the People's Republic of China (Draft for Suggestions from Scholars)* [中华人民共和国人工智能法 (学者建议稿)], trans. by Center for Security and Emerging Technology, March 2024; Zhou Hui, Zhu Yue, Zhu Lingfen, Su Yu, Yao Zhiwei, Wang Jun, Chen Tianhao, He Bo, Hong Yanqing, Fu Hongyu, et al., *The Model Artificial Intelligence Law (MAIL) v.2.0—Multilingual Version* [人工智能示范法2.0], Zenodo, April 2024.

[30] Matt Sheehan, *Tracing the Roots of China's AI Regulations*, Carnegie Endowment for International Peace, February 2024a.

[31] Central Committee of the Communist Party of China, 2024.

[32] *Track I* diplomacy consists of official diplomacy; *Track II* diplomacy refers to unofficial or nongovernmental dialogues (e.g., those sponsored by think tanks).

[33] See Center for International Security and Strategy, "CISS Organizes the Tenth Round of U.S.-China Dialogue on Artificial Intelligence and International Security," July 1, 2024; Fu Ying and John Allen, "Together the U.S. and China Can Reduce the Risks from AI," *Noema*, December 17, 2020; National Committee on U.S.-China Relations, "U.S.-China Track II Dialogue on the Digital Economy," webpage, undated; Meaghan Tobin, "A.I. Pioneers Call for Protections Against 'Catastrophic Risks,'" *New York Times*, September 16, 2024.

[34] Valerie Shen and Jim Kessler, "Competing Values Will Shape US-China AI Race," *Third Way*, July 2024.

A China-led AI resolution at the UN focused significantly on AI capacity-building and was supported by the United States.[35] Nevertheless, which countries should be able to develop and deploy GPAI is likely to be a point of contention between the two countries. National security interests and U.S.-China competition have motivated the United States to restrict China's access to some AI-related hardware. In particular, the United States has restricted exports of the chips that are generally used when training GPAI systems to China and some other countries.[36] The stated goal of these restrictions is to slow Chinese military use of AI, though the restrictions apply generally to the Chinese AI industry, likely because of concerns about links between Chinese companies and China's military.[37]

Unsurprisingly, current hardware export controls have been strongly criticized by China.[38] More generally, China's Global AI Governance Initiative stresses that "[a]ll countries, regardless of their size, strength, or social system, should have equal rights to develop and use AI," and the Shanghai Declaration on Global AI Governance "advocate[s] the spirit of openness and shared benefit."[39]

A specific divergence on capacity-building is that U.S. policymakers might be more concerned than their Chinese counterparts about *open-weight* or *open-source models*, which are models that anyone can use and modify because developers have made their weights publicly available. Key Chinese policy documents, such as the communiqué about the Global AI Governance Initiative, endorse the principle of open-source models with few explicit caveats, which could, in part, be driven by broader economic considerations.[40] Although key U.S. documents highlight the benefits of such models, they also note risks (including risks that do not relate to China). A 2024 report on open models states that restrictions are not currently necessary but could be in the future if, for example, open models substantially lower barriers of entry for non-experts to acquire chemical, biological, radiological, or nuclear weapons.[41]

However, at least some of this divergence could be driven by differences in which kinds of AI systems policymakers have in mind rather than by differences in how policymakers would treat AI systems that they knew could be helpful against chemical, biological, radiological, or nuclear attacks. It would also be unsurprising if Chinese policymakers were to primarily prioritize relatively unsophisticated—thus less capable—AI systems when referring to the rights of all countries to develop and use AI. For example, previous Chinese strategies involving technological exports to the Global South have focused on providing low-cost alternatives to existing data infrastructure rather than funding frontier research.[42]

---

[35] UN General Assembly, 2024b.

[36] Allen, 2022.

[37] Allen, 2022.

[38] "China Lashes Out at Latest U.S. Export Controls on Chips," Associated Press, October 8, 2022.

[39] Ministry of Foreign Affairs of the People's Republic of China, "Shanghai Declaration on Global AI Governance," July 4, 2024a; Ministry of Foreign Affairs of the People's Republic of China, "Global AI Governance Initiative," October 20, 2024c.

[40] Caroline Meinhardt, "Open Source of Trouble: China's Efforts to Decouple from Foreign IT Technologies," Mercator Institute for China Studies, March 18, 2020; Zeyi Yang, "Why Chinese Companies Are Betting on Open-Source AI," *MIT Technology Review*, July 24, 2024.

[41] National Telecommunications and Information Administration, *Dual-Use Foundation Models with Widely Available Model Weights*, July 2024.

[42] Evan Williams, "China's Digital Silk Road Taking Its Shot at the Global Stage," *East Asia Forum*, May 9, 2024.

Control is also a concept for which divergences between the United States and China could be particularly important. Experts in both countries have warned of the potential for human operators to lose control of GPAI systems.[43] However, some Chinese discussions of control have been framed around "negative impact[s] to social values."[44] Chinese discussions of control must also be analyzed against a backdrop of the government's long history of content moderation and ideological management.[45] Thus, it could be valuable for the United States and China to create consensus around the term *control* or at least clarify its meaning in particular use cases. A related issue is that the Chinese term, *anquan*, can be translated as both safety and security, which makes it less clear which kinds of AI risks the authors of Chinese policy documents have in mind.[46]

A final divergence is that the Chinese government might be more strongly in favor of using AI to bolster state power at home and abroad.[47] Chinese policy documents sometimes refer to harmony as a goal for AI governance.[48] Some researchers argue that references to harmony might reflect the Chinese government's emphasis on state autonomy by promoting collective goals rather than individual rights.[49] The U.S. government has also expressed concerns about the Chinese government using AI to increase state control: The United States sanctioned Chinese AI companies that are involved in surveillance and expressed concerns during a Track I dialogue in May 2024 about China's misuse of AI.[50] However, AI systems that are most commonly discussed in the context of surveillance and state control, such as facial recognition systems, are generally narrow rather than GPAI.

## Differences in Approaches to Implementing International AI Governance

At the international level, one key divergence between the two countries is their preferred forum for implementing AI governance; China more strongly favors governance through the UN, while the United States has favored more-exclusive settings. When announcing its flagship Global AI Governance Initiative, for instance, Chinese leadership stated that it supports discussions "within the UN framework" about new international institutions for AI governance.[51] Beijing has subsequently called for the UN to be the main channel for international AI governance, including in its readout of the U.S.-China intergovernmental dialogue and in the Shanghai Declaration.[52] The Chinese government has also engaged with AI governance forums outside the UN, such as the Bletchley Summit and its successor in South Korea, as well as the U.S.-China intergovernmental dialogue.[53]

---

[43] International Dialogues on AI Safety, undated; Sheehan, 2024b.

[44] National Technical Committee 260 on Cybersecurity of Standardization Administration of China, [络安全标准实践指南—人工智能伦理安全风险防范指]引], 2024.

[45] Concordia AI, "China's AI Safety Evaluations Ecosystem," *AI Safety in China* blog, September 13, 2024b; Matt Sheehan, "China's AI Regulations and How They Get Made," Carnegie Endowment for International Peace, July 10, 2023.

[46] Sheehan, 2024b.

[47] Paul Scharre, *The Dangers of the Global Spread of China's Digital Authoritarianism*, Center for a New American Security, May 2023.

[48] Bruce Sterling, "The Beijing Artificial Intelligence Principles," *Wired*, June 1, 2019.

[49] Manoj Kewalramani, "How China's Global Initiatives Aim to Change the Way the World Is Run," *Wire China*, October 22, 2023; Huw Roberts, Josh Cowls, Emmie Hine, Jessica Morley, Vincent Wang, Mariarosaria Taddeo, and Luciano Floridi, "Governing Artificial Intelligence in China and the European Union: Comparing Aims and Promoting Ethical Outcomes," *Information Society*, Vol. 39, No. 2, March 2023.

[50] William Crumpler and William A. Carter, "Understanding the Entities Listing in the Context of U.S.-China AI Competition," Center for Strategic and International Studies, October 15, 2019; White House, 2024.

[51] Ministry of Foreign Affairs of the People's Republic of China, 2024c.

The United States has worked with China on AI governance at the UN. In addition to the discussed resolutions, U.S. and Chinese experts have participated in the UN's high-level AI Advisory Body.[54] Generally, however, the United States seems to oppose efforts for major new institutions or commitments around GPAI at the UN level.[55] The United States has instead favored more-exclusive settings, particularly settings that primarily engage U.S. allies. For instance, the United States has promoted the G7 Hiroshima Process and reportedly expressed concerns about China's participation in the Bletchley Summit.[56] China has also attempted to kickstart international governance efforts with aligned countries, such as the BRICS study group on AI,[57] but it does not seem to be prioritizing these initiatives.[58]

Both countries have also taken differing approaches to diplomacy around AI safety. As other commentators have noted, the U.S. delegation to the U.S.-China AI dialogue was led by national security authorities, whereas the Chinese delegation was led by officials from the Ministry of Foreign Affairs—specifically, the department that handles relations with the United States.[59] This divergence might suggest that the United States was more focused on safety, security, and ensuring that AI would not be misused; the Chinese side might have been more concerned with China's international image and managing the bilateral relationship.

Similarly, the United States has established an AI safety institute, which is now part of an international network of such institutions that primarily links countries that are allied to the United States.[60] At the time of this writing, the Chinese government has not (yet) created an AI safety institute.[61] That said, it intends to "institute oversight systems to ensure the safety of artificial intelligence," and there are already institutions in China that are somewhat analogous.[62] For example, the China Academy of Information and Communications Technology, a government-linked think tank, runs evaluations on Chinese models, including for some types of safety.[63] China was not invited to be a member of the U.S.-led network of AI safety institutes; however, whether a Chinese institution is eventually included might be an important indicator of U.S. attitudes toward cooperation with China around mitigating AI risks.

---

[52] "China Calls on the International Community to Work Together to Promote Global Artificial Intelligence Governance" ["中方呼吁国际社会携手合作共同推进全球人工智能治理"], Xinhua, May 9, 2024; "China's Readout of the First Sino-U.S. Intergovernmental Dialogue on AI," 2024; Ministry of Foreign Affairs of the People's Republic of China, 2024a.

[53] Government of the United Kingdom, 2023; Government of the United Kingdom, 2024b.

[54] UN, "AI Advisory Body," webpage, undated.

[55] Fiona Alexander, "UN Attempts AI Power Grab. The West Is Unhappy," Center for European Policy Analysis, July 24, 2024; Colum Lynch, "Biden Administration Douses UN's AI Governance Aspirations," *Devex*, April 25, 2024.

[56] Government of Japan, "The Hiroshima AI Process: Leading the Global Challenge to Shape Inclusive Governance for Generative AI," February 9, 2024; Vincent Manancourt, Tom Bristow, and Laurie Clarke, "China Expected at UK AI Summit Despite Pushback from Allies," *Politico*, August 25, 2023.

[57] BRICS is an intergovernmental organization comprising Brazil, Russia, India, China, South Africa, Iran, Egypt, Ethiopia, and the United Arab Emirates.

[58] Xi Jinping, speech delivered at the 15th BRICS Leaders' Meeting, August 23, 2023.

[59] "A Few Thoughts on the First Sino-US Intergovernmental Dialogue on AI," *Geopolitechs* blog, May 17, 2024.

[60] U.S. Department of Commerce, "U.S. Secretary of Commerce Gina Raimondo Releases Strategic Vision on AI Safety, Announces Plan for Global Cooperation Among AI Safety Institutes," press release, May 21, 2024.

[61] Lily Ottinger, "Where's China's AI Safety Institute?" *ChinaTalk* blog, November 20, 2024.

[62] Central Committee of the Communist Party of China, 2024; Karson Elmgren and Oliver Guest, *Chinese AISI Counterparts*, Institute for AI Policy and Strategy, October 2024.

## Toward U.S.-China AI Safety Cooperation

Overall, further dialogue and consensus-building might be helpful for reducing divergences in approaches to AI governance and making progress on cooperation to reduce safety and security risks. Discussions could happen in both the official U.S.-China Track I dialogue on AI, in various Track II processes, and through other back channels.[64] Where differences remain, policymakers can take action to facilitate cooperation on mitigating these risks.

First, policymakers could focus on areas in which the United States and China most clearly have aligned interests, such as addressing bioweapon proliferation risks from malicious nonstate actors that are using AI tools.[65] Narrow cooperation could help combat the highest-severity risks and build trust for more-expansive efforts, though such initiatives must be carefully considered given the rapid pace of AI development and the opportunity costs for both countries.

Second, when designing international AI governance proposals, policymakers could aim for approaches that are interoperable or harmonized across jurisdictions. For example, international governance mechanisms could address both AI models and specific kinds of AI-generated content, such as outputs that could contribute to bioweapons proliferation.

Third, given that key terms are sometimes understood differently in the two countries, it might be valuable to ensure that key terms, such as *control*, are clearly defined in shared statements and international agreements (although constructive ambiguity might occasionally be helpful for reaching agreement).[66]

Researchers and policymakers in the United States and China have expressed concerns about safety and security risks from GPAI, and both sides have demonstrated a willingness to govern such systems. Although there are still many barriers to cooperation, such as national security considerations and the delicate U.S.-China relationship, U.S. and Chinese policymakers can work toward collaboration on reducing these risks by better understanding how each views AI governance.

---

[63] Jeffrey Ding, "CAICT's 7th Batch of AI Model Evaluations," *ChinAI Newsletter* blog, August 5, 2024.

[64] Lucero, 2024.

[65] David Heslop and Joel Keep, "The 2024 China-US AI Dialogue Should Start with an Eye on Chem-Bio Weapons," *The Diplomat*, March 9, 2024.

[66] Michael Byers, "Still Agreeing to Disagree: International Security and Constructive Ambiguity," *Journal on the Use of Force and International Law*, Vol. 8, No. 1, January 2021.

# References

"A Few Thoughts on the First Sino-US Intergovernmental Dialogue on AI," *Geopolitechs* blog, May 17, 2024.

Alexander, Fiona, "UN Attempts AI Power Grab. The West Is Unhappy," Center for European Policy Analysis, July 24, 2024.

Allen, Gregory C., "Choking Off China's Access to the Future of AI," Center for Strategic and International Studies, October 11, 2022.

Bengio, Y., Mindermann, S., Privitera, D., Besiroglu, T., Bommasani, R., Casper, S., Choi, Y., Goldfarb, D., Heidari, H., Khalatbari, L., et al., *International Scientific Report on the Science of AI: Interim Report*, Government of the United Kingdom, May 2024.

Bengio, Yoshua, Geoffrey Hinton, Andrew Yao, Dawn Song, Pieter Abbeel, Trevor Darrell, Yuval Noah Harari, Ya-Qin Zhang, Lan Xue, Shai Shalev-Shwartz, et al., "Managing Extreme AI Risks Amid Rapid Progress," *Science*, May 20, 2024.

Beraja, Martin, Andrew Kao, David Y. Yang, and Noam Yuchtman, "Exporting the Surveillance State via Trade in AI," Brookings Institution, July 24, 2023.

Beraja, Martin, David Y. Yang, and Noam Yuchtman, "China Is Exporting Its AI Surveillance State," *Project Syndicate*, July 24, 2024.

Browne, Ryan, "Google Deepmind Boss Hits Back at Meta AI Chief over 'Fearmongering' Claim," CNBC, October 31, 2023.

Brundage, Miles, Shahar Avin, Jasmine Wang, Haydn Belfield, Gretchen Krueger, Gillian Hadfield, Heidy Khlaaf, Jingying Yang, Helen Toner, Ruth Fong, et al., "Toward Trustworthy AI Development: Mechanisms for Supporting Verifiable Claims," arXiv, arXiv:2004.07213, April 20, 2020.

Byers, Michael, "Still Agreeing to Disagree: International Security and Constructive Ambiguity," *Journal on the Use of Force and International Law*, Vol. 8, No. 1, January 2021.

California State Senate, Safe and Secure Innovation for Frontier Artificial Intelligence Models Act, Senate Bill 1047, September 2, 2024.

Cave, Stephen, Seán S. ÓhÉigeartaigh, "An AI Race for Strategic Advantage: Rhetoric and Risks," *AIES '18: Proceedings of the 2018 AAAI/ACM Conference on AI, Ethics, and Society*, December 2018.

Center for International Security and Strategy, "CISS Organizes the Tenth Round of U.S.-China Dialogue on Artificial Intelligence and International Security," July 1, 2024.

Central Committee of the Communist Party of China, resolution on Further Deepening Reform Comprehensively to Advance Chinese Modernization, July 18, 2024.

Chandra Mishra, Nayan, "From Competition to Cooperation: Can US-China Engagement Overcome Geopolitical Barriers in AI Governance?" *Tech Policy Press*, September 23, 2024.

"China and the U.S. Agreed to Hold the 2nd Intergovernmental Dialogue on AI," *Geopolitechs* blog, August 28, 2024.

"China Calls on the International Community to Work Together to Promote Global Artificial Intelligence Governance" ["中方呼吁国际社会携手合作共同推进全球人工智能治理"], Xinhua, May 9, 2024. As of December 4, 2024: https://www.qinshui.gov.cn/xwdt_361/gwyyw_49700/202405/t20240509_1979919.shtml

"China Lashes Out at Latest U.S. Export Controls on Chips," Associated Press, October 8, 2022.

"China's Readout of the First Sino-U.S. Intergovernmental Dialogue on AI," *Geopolitechs* blog, May 14, 2024.

Concordia AI, *State of AI Safety in China*, October 2023.

Concordia AI, "What Does the Chinese Leadership Mean by 'Instituting Oversight Systems to Ensure the Safety of AI?'" *AI Safety in China* blog, August 2, 2024a.

Concordia AI, "China's AI Safety Evaluations Ecosystem," *AI Safety in China* blog, September 13, 2024b.

Cottier, Ben, Tamay Besiroglu, and David Owen, "Who Is Leading in AI? An Analysis of Industry AI Research," arXiv, arXiv:2312.00043, November 24, 2023.

Crumpler, William, and William A. Carter, "Understanding the Entities Listing in the Context of U.S.-China AI Competition," Center for Strategic and International Studies, October 15, 2019.

Cuéllar, Mariano-Florentino, and Matt Sheehan, "AI Is Winning the AI Race," *Foreign Policy*, June 19, 2023.

Cyberspace Administration of China, "Regulations on the In-Depth Synthesis Management of Internet Information Services" ["互联网信息服务深度合成管理规定"], December 11, 2022. As of December 4, 2024: https://www.cac.gov.cn/2022-12/11/c_1672221949354811.htm

Cyberspace Administration of China, "Interim Measures for the Management of Generative Artificial Intelligence Services" ["生成式人工智能服务管理暂行办法"], July 13, 2023. As of December 4, 2024: https://www.cac.gov.cn/2023-07/13/c_1690898327029107.htm

Ding, Jeffrey, "CAICT's 7th Batch of AI Model Evaluations," *ChinAI Newsletter* blog, August 5, 2024.

Elmgren, Karson, and Oliver Guest, *Chinese AISI Counterparts*, Institute for AI Policy and Strategy, October 2024.

Executive Order 14110, "Safe, Secure, and Trustworthy Development and Use of Artificial Intelligence," Executive Office of the President, October 30, 2023.

Fu Ying and John Allen, "Together the U.S. and China Can Reduce the Risks from AI," *Noema*, December 17, 2020.

General Assembly of the State of Colorado, Concerning Consumer Protections in Interactions with Artificial Intelligence Systems, Senate Bill 24-205, May 17, 2024.

Government of Japan, "The Hiroshima AI Process: Leading the Global Challenge to Shape Inclusive Governance for Generative AI," February 9, 2024.

Government of the United Kingdom, "The Bletchley Declaration by Countries Attending the AI Safety Summit, 1–2 November 2023," policy paper, November 1, 2023.

Government of the United Kingdom, "Historic First as Companies Spanning North America, Asia, Europe and Middle East Agree Safety Commitments on Development of AI," press release, May 21, 2024a.

Government of the United Kingdom, "AI Seoul Summit: Participants List (Governments and Organisations)," webpage, last updated June 13, 2024b. As of November 26, 2024:
https://www.gov.uk/government/publications/ai-seoul-summit-programme/ai-seoul-summit-participants-list-governments-and-organisations

Heim, Lennart, and Leonie Koessler, "Training Compute Thresholds: Features and Functions in AI Regulation," arXiv, arXiv:2405.10799, August 6, 2024.

Heslop, David, and Joel Keep, "The 2024 China-US AI Dialogue Should Start with an Eye on Chem-Bio Weapons," *The Diplomat*, March 9, 2024.

International Dialogues on AI Safety, homepage, undated. As of November 25, 2024:
https://idais.ai/

International Dialogues on AI Safety, statement at IDAIS-Beijing 2024, March 10–11, 2024. As of December 4, 2024:
https://idais.ai/dialogue/idais-beijing/

"Is Xi Jinping an AI Doomer?" *The Economist*, August 25, 2024.

Kewalramani, Manoj, "How China's Global Initiatives Aim to Change the Way the World Is Run," *Wire China*, October 22, 2023.

Kissinger, Henry A., and Graham Allison, "The Path to AI Arms Control," *Foreign Affairs*, October 13, 2023.

Liu, Qianer, "China to Lay Down AI Rules with Emphasis on Content Control," *Financial Times*, July 10, 2023.

Lucero, Karman, "Managing the Sino-American AI Race," *Project Syndicate*, August 9, 2024.

Lynch, Colum, "Biden Administration Douses UN's AI Governance Aspirations," *Devex*, April 25, 2024.

Manancourt, Vincent, Tom Bristow, and Laurie Clarke, "China Expected at UK AI Summit Despite Pushback from Allies," *Politico*, August 25, 2023.

Meinhardt, Caroline, "Open Source of Trouble: China's Efforts to Decouple from Foreign IT Technologies," Mercator Institute for China Studies, March 18, 2020.

Ministry of Foreign Affairs of the People's Republic of China, "Shanghai Declaration on Global AI Governance," July 4, 2024a. As of December 4, 2024:
https://www.mfa.gov.cn/eng/xw/zyxw/202407/t20240704_11448351.html

Ministry of Foreign Affairs of the People's Republic of China, "AI Capacity-Building Action Plan for Good and for All," September 27, 2024b. As of December 4, 2024:
https://www.mfa.gov.cn/eng/wjbzhd/202409/t20240927_11498465.html

Ministry of Foreign Affairs of the People's Republic of China, "Global AI Governance Initiative," October 20, 2024c. As of December 4, 2024:
https://www.mfa.gov.cn/eng/zy/gb/202405/t20240531_11367503.html

Murgia, Madhumita, "White House Science Chief Signals US-China Co-Operation on AI Safety," *Financial Times*, January 24, 2024.

National Committee on U.S.-China Relations, "U.S.-China Track II Dialogue on the Digital Economy," webpage, undated. As of November 26, 2024:
https://www.ncuscr.org/program/us-china-track-ii-dialogue-digital-economy/

National Technical Committee 260 on Cybersecurity of Standardization Administration of China, *Basic Safety Requirements for Generative Artificial Intelligence Services* [生成式人工智能服务安全基本要求], trans. by Center for Security and Emerging Technology, February 2024. As of December 4, 2024:
https://cset.georgetown.edu/publication/china-safety-requirements-for-generative-ai-final/

National Technical Committee 260 on Cybersecurity of Standardization Administration of China, 网络安全标准实践指南—人工智能伦理安全风险防范指引, January 2021. As of December 4, 2024:
https://www.tc260.org.cn/file/zn10.pdf

National Telecommunications and Information Administration, *Dual-Use Foundation Models with Widely Available Model Weights*, July 2024.

Nye, Joseph S., Jr., "U.S.–China Cooperation Remains Possible," *Project Syndicate*, May 6, 2024.

Ottinger, Lily, "Where's China's AI Safety Institute?" *ChinaTalk* blog, November 20, 2024.

Roberts, Huw, "China's Ambitions for Global AI Governance," *East Asia Forum*, September 10, 2024.

Roberts, Huw, Josh Cowls, Emmie Hine, Jessica Morley, Vincent Wang, Mariarosaria Taddeo, and Luciano Floridi, "Governing Artificial Intelligence in China and the European Union: Comparing Aims and Promoting Ethical Outcomes," *Information Society*, Vol. 39, No. 2, March 2023.

"Sabotage Evaluations for Frontier Models," *Anthropic*, October 18, 2024.

Scharre, Paul, *The Dangers of the Global Spread of China's Digital Authoritarianism*, Center for a New American Security, May 2023.

Sharkey, Lee, Clíodhna Ní Ghuidhir, Dan Braun, Jérémy Sheurer, Mikita Balesni, Lucius Bushnaq, Charlotte Stix, and Marius Hobbhahn, *A Causal Framework for AI Regulation and Auditing*, Apollo Research, November 2024.

Shavit, Yonadav, "What Does It Take to Catch a Chinchilla? Verifying Rules on Large-Scale Neural Network Training via Compute Monitoring," arXiv, arXiv: 2303.1134, May 30, 2023.

Sheehan, Matt, "China's AI Regulations and How They Get Made," Carnegie Endowment for International Peace, July 10, 2023.

Sheehan, Matt, *Tracing the Roots of China's AI Regulations*, Carnegie Endowment for International Peace, February 2024a.

Sheehan, Matt, "China's Views on AI Safety Are Changing—Quickly," Carnegie Endowment for International Peace, August 27, 2024b.

Shen, Valerie, and Jim Kessler, "Competing Values Will Shape US-China AI Race," *Third Way*, July 2024.

Sihao Huang, "Beijing's Vision of AI Global Governance," *ChinaTalk* blog, October 23, 2023.

Sterling, Bruce, "The Beijing Artificial Intelligence Principles," *Wired*, June 1, 2019.

Sun Chenghao and Lie Yuan, "How Should China and the United States Promote Artificial Intelligence Dialogue and Cooperation?" ["中美应如何推动人工智能对话与合作"], *China US Focus*, February 6, 2024.

Thorbecke, Catherine, "Where the U.S. and China Can Find Common Ground on AI," Bloomberg, September 19, 2024.

Thornhill, John, "Step Aside World, the U.S. Wants to Write the AI Rules," *Financial Times*, November 2, 2023.

Tobin, Meaghan, "A.I. Pioneers Call for Protections Against 'Catastrophic Risks,'" *New York Times*, September 16, 2024.

UN—*See* United Nations.

United Nations, "AI Advisory Body," webpage, undated. As of November 27, 2024: https://www.un.org/ai-advisory-body

United Nations General Assembly, resolution on Seizing the Opportunities of Safe, Secure and Trustworthy Artificial Intelligence Systems for Sustainable Development, Resolution A/78/L.49, March 11, 2024a.

United Nations General Assembly, Resolution on Enhancing International Cooperation on Capacity-Building of Artificial Intelligence, Resolution A/78/L.86, June 25, 2024b.

U.S. Department of Commerce, "U.S. Secretary of Commerce Gina Raimondo Releases Strategic Vision on AI Safety, Announces Plan for Global Cooperation Among AI Safety Institutes," press release, May 21, 2024.

Webster, Graham, "What If 'AI Safety' Is No Longer a Constructive U.S.-China Topic?" *Here It Comes* blog, August 23, 2024.

Weitz, Richard, "China and the United States Begin Official AI Dialogue," *China US Focus*, June 14, 2024.

White House, "Statement from NSC Spokesperson Adrienne Watson on the U.S.-PRC Talks on AI Risk and Safety," May 15, 2024.

Williams, Evan, "China's Digital Silk Road Taking Its Shot at the Global Stage," *East Asia Forum*, May 9, 2024.

Xi Jinping, speech delivered at the 15th BRICS Leaders' Meeting, August 23, 2023.

Yang, Zeyi, "Why Chinese Companies Are Betting on Open-Source AI," *MIT Technology Review*, July 24, 2024.

Zhang Linghan, Yang Jianjun, Cheng Ying, Zhao Jingwu, Han Xuzhi, Zheng Zhifeng, and Xu Xiaob, *Artificial Intelligence Law of the People's Republic of China (Draft for Suggestions from Scholars)* [中华人民共和国人工智能法 (学者建议稿)], trans. by Center for Security and Emerging Technology, March 2024. As of December 4, 2024: https://cset.georgetown.edu/publication/china-ai-law-draft/

Zhou Hui, Zhu Yue, Zhu Lingfen, Su Yu, Yao Zhiwei, Wang Jun, Chen Tianhao, He Bo, Hong Yanqing, Fu Hongyu, et al., *The Model Artificial Intelligence Law (MAIL) v.2.0—Multilingual Version* [人工智能示范法2.0], Zenodo, April 2024. As of December 4, 2024: https://zenodo.org/records/10974163

# About the Authors

**Oliver Guest** is a research analyst in the international governance team at the Institute for AI Policy and Strategy. His research focuses on AI in China, coordination between AI safety institutes, and AI's geopolitical effects. He holds an M.A. in security studies.

**Kevin Wei** is a 2023 Schwarzman Scholar and a Technology and Security Policy fellow at RAND (for more information on the fellowship program, visit www.rand.org/tasp-fellows). Wei conducts legal and technical research on the governance of advanced AI systems. They have an M.S. in machine learning and an M.A. in global affairs.